\documentclass{article}
\usepackage{amsfonts}
\usepackage{graphicx}
\usepackage{epsfig}
\usepackage{eepic}
\usepackage{cite}

\newcommand{\nc}{\newcommand}
\nc{\rnc}{\renewcommand}
\nc{\ket}[1]{\left | \, #1 \right \rangle}
\nc{\bra}[1]{\left \langle #1 \, \right |}
\nc{\proj}[1]{\ket{#1}\bra{#1}}
\rnc{\vec}{\mathbf}
\nc{\ua}{\uparrow}
\nc{\da}{\downarrow}

\nc{\braket}[2]{\langle\, #1\,|\,#2\,\rangle}
\nc{\half}{\frac{1}{2}}

\nc{\prj}{\mathcal{P}}
\nc{\hilb}{\mathcal{H}}
\nc{\pth}{\mathcal{C}}
\nc{\inprod}[2]{\braket{#1}{#2}}
\nc{\upket}{\ket{\uparrow}}
\nc{\downket}{\ket{\downarrow}}
\nc{\upbra}{\bra{\uparrow}}
\nc{\downbra}{\bra{\downarrow}}

\def\Id{{\mathbf 1}}
\def\d{{\rm d}}

\begin{document}

\title{Geometric Quantum Computation}
\author{
Artur Ekert \and
Marie Ericsson\thanks{Department of Quantum Chemistry, Uppsala University,
Box 518, Se-751 20 Uppsala, Sweden} \and
Patrick Hayden \and
Hitoshi Inamori \and
Jonathan A. Jones \and
Daniel K. L. Oi \and
Vlatko Vedral \\
\\
Centre for Quantum Computation \\
University of Oxford \\
Clarendon Laboratory, Parks Road \\
Oxford OX1 3PU, UK}
\maketitle

\begin{abstract}
We describe in detail a general strategy for implementing a
conditional geometric phase between two spins.
Combined with single-spin operations, this simple operation is a
universal gate for quantum computation, in that any unitary
transformation can be implemented with arbitrary precision using
only single-spin operations and conditional phase shifts. Thus quantum
geometrical phases can form the basis of any quantum computation.
Moreover, as the induced conditional phase depends only on the
geometry of the paths executed by the spins it is resilient
to certain types of errors and offers the potential of a naturally
fault-tolerant way of performing quantum computation.
\end{abstract}

\section{Introduction} \label{sIntro}
Among the surprising effects recently discovered in quantum
mechanics is that a quantum system retains a memory of its motion when
it undergoes a cyclic evolution \cite{Ber84}. This is
reflected in the existence of the Berry phase, a phase acquired by the
quantum state of the system in addition to the better known dynamic
phase. The Berry phase is a purely geometrical effect that can be
linked to the notion of parallel transport \cite{BS83}: it
depends only on the area covered by the motion of the system, and is
independent of details of how the motion is executed. Berry phases
have been demonstrated in a wide variety of systems \cite{GPP89},
including NMR \cite{DS87,GFG96}, the closely related technique of
NQR \cite{RT87,MM94,JAJ97}, optical systems \cite{Tomita87}, and
others.

An equally exciting recent development in the field of quantum
mechanics has been the discovery that quantum systems can be used
to perform novel information processing tasks, including
computations which are more efficient than any algorithm known on
a classical computer~\cite{Deu85,Shor94,Gro96}. Quantum
information processing requires the ability to execute conditional
dynamics \cite{cqdlg95} between two quantum bits (qubits), where
the state of one qubit influences the evolution of another qubit
during a quantum computation.  Simple quantum information
processing has been demonstrated using NMR \cite{Cory96, Cory97,
Gersh97, Jones98} and trapped ions \cite{Monroe95}.

Recent experimental work has managed to combine these two
quantum phenomena in the form of geometric quantum
computation~\cite{Jones00}.  (For a more abstract approach see
\cite{Zan99,Pac99}.)
In this paper we seek to detail the theoretical
ideas behind geometric quantum computation.
In particular we show that Berry's phase may be used to
implement conditional phase shifts, and thus any quantum
gate~\cite{BBC95}.
We begin with brief introductions to both quantum gates and
networks as well as to geometric phases, proceeding to analyse the
dynamics of a spin-half system in order to see in detail how the
theory of geometric phases applies there.  Finally, we
extend the ideas to pairs of spin-half particles,
showing how to introduce a conditional geometric phase between
the two particles.

\section{Phase gates and quantum computation} \label{sBasics}

\subsection{Qubits and networks}
A \emph{qubit} is a quantum system in which the Boolean states $0$
and $1$ are represented by a prescribed pair of normalised and
mutually orthogonal quantum states labeled as $\{|0\rangle
,|1\rangle \}$.  Unlike a simple Boolean variable, a
qubit, typically a microscopic system such as an atom, a nuclear
spin, or  a polarised photon, can exist in an arbitrary
superposition $\alpha \ket{0} + \beta \ket{1}$, making it more
powerful as a computational resource.

In quantum computation, we set some \emph{register} of qubits to
an ``input'' state, evolve the qubits unitarily using simple
building-block operations and then take the final state as
``output''. More formally, a \emph{quantum logic gate} is a device
which performs a fixed unitary operation on selected qubits in a
fixed period of time and a \emph{quantum network} is a device
consisting of quantum logic gates whose computational steps are
synchronised in time~\cite{Deu89}. The outputs of some of the
gates are connected by wires to the inputs of others. The
\emph{size} of the network is the number of gates it contains.

\subsection{Quantum logic gates}
The most common quantum gate is the Hadamard gate, a single
qubit gate $H$ performing the unitary transformation known as the
Hadamard transform.  It is defined as
\setlength{\unitlength}{0.030in}
\begin{equation}
H=\frac{1}{\sqrt{2}}\left(
\begin{array}{cc}
1 & 1 \\
1 & -1
\end{array}
\right) \mbox{\hspace{2cm}} \mbox{
\begin{picture}(60,0)(15,15)
  \put(-4,14){$\ket{x}$} \put(5,15){\line(1,0){5}} \put(20,15){\line(1,0){5}}
  \put(10,10){\framebox(10,10){$H$}}
\put(30,14){$\displaystyle\frac{(-1)^x\ket{x}+\ket{1-x}}{\sqrt{2}}$}
\end{picture}
}
\end{equation}
The matrix is written in the computational basis
$\{\ket{0}, \ket{1} \}$
and the diagram on
the right provides a schematic representation of the gate $H$
acting on a qubit in state $\ket{x}$, with $x=0,1$.

The addition of another single qubit gate, the phase shift gate
$\mathbf{\phi }$, defined as $\left|
\,0\right\rangle \mapsto \left| \,0\right\rangle $ and $\left|
\,1\right\rangle \mapsto e^{i\phi }\left| \,1\right\rangle $, or,
in matrix notation,
\setlength{\unitlength}{0.030in}
\begin{equation}
{\mathbf{\phi}} = \left (
\begin{array}{cc}
1 & 0 \\
0 & e^{i\phi}
\end{array}
\right ) \mbox{\hspace{3cm}}
\mbox{
\begin{picture}(30,0)(15,15)
  \put(-4,14){$\ket{x}$} \put(5,15){\line(1,0){20}} \put(20,15){\line(1,0){5}}
  \put(15,15){\circle*{3}}
\put(14,19){$\phi $}
\put(30,14){$e^{ix\phi}\ket{x}$}
\end{picture}
}
\end{equation}
is actually sufficient to construct
the following network (of size four), which generates the most
general pure state of a single qubit (up to a global phase),
\setlength{\unitlength}{0.025in}
\begin{equation}\label{1quniv}
\mbox{
\begin{picture}(80,12)(0,3)

\put(0,5){$\ket{0}$}

\put(10,5){\line(1,0){10}}
\put(20,0){\framebox(10,10){$H$}}
\put(30,5){\line(1,0){20}}
\put(50,0){\framebox(10,10){$H$}}
\put(60,5){\line(1,0){20}}

\put(40,5){\circle*{3}}
\put(38,10){$2\theta$}

\put(70,5){\circle*{3}}
\put(65,10){$\frac{\pi}{2}+\phi$}

\end{picture}
}
\quad
\cos\theta\ket{0}+e^{i\phi}\sin\theta\ket{1}.
\end{equation}
Consequently, the Hadamard and phase gates are sufficient to construct
\emph{any} unitary operation on a single qubit.

Thus the Hadamard gates and the phase gates can be used to
transform the input state $|0\rangle |0\rangle ...|0\rangle $ of
$n$ qubits into any state of the type $|\Psi
_{1}\rangle $ $|\Psi _{2}\rangle ...$ $ |\Psi _{n}\rangle ,$ where
$|\Psi _{i}\rangle $ is an arbitrary superposition of $|0\rangle $
and $|1\rangle .$ These are rather special $n$-qubit states, called
the product states or the separable states. In general, a register of
$n$ qubits can be prepared in states which are not
separable, known as entangled states.

However, in order to
entangle two or more qubits it is necessary to have access
to two-qubit gates.
One such gate is the controlled phase shift gate
$B(\phi)$ defined as
\begin{equation}
{B}(\phi )=\left. \left(
\begin{array}{cccc}
1 & 0 & 0 & 0 \\
0 & 1 & 0 & 0 \\
0 & 0 & 1 & 0 \\
0 & 0 & 0 & e^{i\phi }
\end{array}
\right) \mbox{\hspace{1.5cm}}
\mbox{
\begin{picture}(25,0)(0,20)
  \put(-4,14){$\ket{y}$} \put(-4,29){$\ket{x}$} \put(5,15){\line(1,0){20}}
\put(5,30){\line(1,0){20}}
  \put(15,30){\circle*{3}} \put(15,15){\line(0,1){15}} \put(17,21){$\phi$}
\put(15,15){\circle*{3}}
\end{picture}}\quad \right\} e^{ixy\phi }\left| \,x\right\rangle \left|
\,y\right\rangle .
\end{equation}
The matrix is written in the computational basis
$\{ \ket{00}, \ket{01}, \ket{10},$ $\ket{11} \}$
and the diagram on the right shows the structure of the gate.

\subsection{Universality}
An important result in the theory of quantum computation states
that the Hadamard gate, and all $B(\phi)$ controlled phase gates
form a {\em universal set of gates}: if the Hadamard gate as well
as all $B(\phi)$ gates are available then any $n$-qubit unitary
operation can be simulated exactly with less than $C4^{n}n$ such
gates, for some constant $C$~\cite{BBC95}.  Consequently, being
able to implement 1 and 2-qubit phase gates is of crucial
importance in quantum computation.  In this paper we describe a
new method for implementing the controlled phase gates based
explicitly on geometric phases~\cite{Ber84, BS83, GPP89} rather
than dynamic ones.

\section{Geometric phase} \label{sPhase}

\subsection{Cyclic evolution}
The states of a quantum system are usually described as being
represented by vectors of norm 1 ($|\inprod{\psi}{\psi}|^2=1$)
in a complex Hilbert space $\hilb$.  However, there is redundancy
in this description since the state $\ket{\psi}$ is physically
indistinguishable from the state $e^{i\phi} \ket{\psi}$.
It is therefore convenient to consider instead the 
projective space $\prj$,
in which vectors are grouped into equivalence classes under the relation
$\ket{\psi} \sim r e^{i\phi} \ket{\psi}$ for any $r>0$ and real $\phi$,
thereby eliminating the ambiguity.
The associated projection
map is
\begin{equation}
\begin{array}{cccl}
\Pi : & \hilb & \rightarrow & \prj \\
& \ket{\psi}  & \mapsto     & [ \ket{\psi} ] =
                              \left\{\ket{\psi'}:
                                \ket{\psi'}=r e^{i\phi}\ket{\psi}\right\}.
\end{array}
\end{equation}

\begin{figure}
\begin{center}
\setlength{\unitlength}{0.00063333in}
\begingroup\makeatletter\ifx\SetFigFont\undefined%
\gdef\SetFigFont#1#2#3#4#5{%
  \reset@font\fontsize{#1}{#2pt}%
  \fontfamily{#3}\fontseries{#4}\fontshape{#5}%
  \selectfont}%
\fi\endgroup%
{
\begin{picture}(4089,6041)(0,-10)
\thicklines
\put(2057,1814){\ellipse{3600}{3600}}
\put(2057,1787){\ellipse{3570}{680}}
\thinlines
\put(2052,3329){\ellipse{1852}{248}}
\path(2042,1829)(4077,5084)
\blacken\path(4038.824,4966.346)(4077.000,5084.000)(3987.948,4998.152)(4038.824,4966.346)
\path(2057,1844)(12,5092)
\blacken\path(101.324,5006.436)(12.000,5092.000)(50.550,4974.467)(101.324,5006.436)
\path(2057,1829)(2057,6014)
\blacken\path(2087.000,5894.000)(2057.000,6014.000)(2027.000,5894.000)(2087.000,5894.000)
\path(3687,4199)(3158,3395)
\blacken\path(3198.897,3511.737)(3158.000,3395.000)(3249.020,3478.757)(3198.897,3511.737)
\path(2116,5114)(2266,5114)
\path(2116,4677)(2265,4674)
\path(2187,5118)(2187,4679)
\blacken\path(1217,4154)(1357,4099)(1222,4064)
\path(1217,4154)(1357,4099)(1222,4064)
\blacken\path(1590,3264)(1730,3209)(1595,3174)
\path(1590,3264)(1730,3209)(1595,3174)
\blacken\path(2367,3269)(2507,3214)(2372,3179)
\path(2367,3269)(2507,3214)(2372,3179)
\blacken\path(2492,3394)(2352,3449)(2487,3484)
\path(2492,3394)(2352,3449)(2487,3484)
\blacken\path(1737,3379)(1597,3434)(1732,3469)
\path(1737,3379)(1597,3434)(1732,3469)
\blacken\path(1332,4924)(1189,4958)(1317,5024)
\path(1332,4924)(1189,4958)(1317,5024)
\blacken\path(2784,4489)(2659,4569)(2804,4589)
\path(2784,4489)(2659,4569)(2804,4589)
\blacken\path(2657,4154)(2792,4109)(2667,4054)
\path(2657,4154)(2792,4109)(2667,4054)
\path(2047,5109)(2045,5109)(2040,5108)
	(2031,5107)(2018,5104)(1998,5101)
	(1973,5097)(1943,5092)(1907,5087)
	(1866,5080)(1821,5072)(1774,5064)
	(1724,5056)(1673,5047)(1622,5038)
	(1572,5029)(1523,5020)(1475,5011)
	(1429,5003)(1385,4994)(1343,4986)
	(1303,4977)(1265,4969)(1230,4961)
	(1195,4952)(1163,4944)(1131,4936)
	(1101,4927)(1072,4919)(1043,4910)
	(1015,4901)(988,4891)(957,4880)
	(926,4869)(895,4857)(864,4844)
	(834,4831)(804,4818)(774,4803)
	(744,4789)(715,4774)(687,4758)
	(659,4742)(632,4725)(607,4709)
	(582,4692)(559,4675)(538,4658)
	(518,4640)(500,4623)(484,4607)
	(469,4590)(457,4574)(447,4557)
	(438,4542)(432,4526)(427,4511)
	(425,4496)(425,4482)(426,4467)
	(430,4453)(435,4439)(443,4425)
	(453,4411)(464,4397)(478,4383)
	(494,4369)(512,4355)(533,4341)
	(555,4327)(579,4313)(605,4299)
	(633,4285)(663,4272)(694,4259)
	(727,4246)(760,4234)(795,4222)
	(830,4211)(866,4200)(902,4189)
	(939,4180)(977,4170)(1014,4161)
	(1052,4153)(1090,4145)(1128,4138)
	(1166,4131)(1196,4125)(1227,4120)
	(1258,4116)(1290,4111)(1323,4106)
	(1356,4102)(1390,4098)(1424,4094)
	(1460,4090)(1496,4087)(1533,4083)
	(1571,4080)(1610,4078)(1649,4075)
	(1689,4073)(1729,4070)(1770,4069)
	(1811,4067)(1853,4066)(1894,4065)
	(1936,4064)(1978,4063)(2020,4063)
	(2061,4063)(2103,4064)(2144,4064)
	(2185,4065)(2226,4066)(2266,4068)
	(2307,4069)(2346,4071)(2386,4073)
	(2425,4076)(2464,4078)(2504,4081)
	(2543,4084)(2587,4088)(2632,4092)
	(2677,4096)(2722,4101)(2768,4106)
	(2814,4112)(2860,4117)(2907,4124)
	(2953,4130)(3000,4137)(3045,4144)
	(3091,4152)(3135,4159)(3179,4168)
	(3221,4176)(3262,4184)(3301,4193)
	(3338,4202)(3374,4211)(3407,4220)
	(3437,4229)(3465,4238)(3490,4247)
	(3513,4256)(3533,4265)(3550,4274)
	(3564,4283)(3575,4291)(3583,4300)
	(3588,4309)(3590,4317)(3590,4326)
	(3586,4334)(3579,4343)(3569,4351)
	(3556,4360)(3541,4368)(3522,4377)
	(3501,4386)(3477,4395)(3451,4403)
	(3423,4412)(3392,4421)(3359,4430)
	(3324,4439)(3287,4448)(3249,4457)
	(3209,4466)(3169,4474)(3128,4482)
	(3086,4491)(3044,4499)(3002,4506)
	(2960,4514)(2919,4521)(2878,4528)
	(2839,4535)(2800,4541)(2762,4547)
	(2726,4553)(2690,4559)(2656,4564)
	(2624,4569)(2593,4574)(2550,4581)
	(2510,4587)(2473,4593)(2437,4598)
	(2404,4604)(2371,4609)(2340,4614)
	(2309,4619)(2279,4624)(2249,4628)
	(2219,4633)(2190,4638)(2162,4642)
	(2137,4646)(2113,4650)(2094,4653)
	(2078,4656)(2066,4658)(2058,4659)
	(2054,4660)(2052,4660)
\put(2327,4814){\makebox(0,0)[lb]{$e^{i\phi}$}}
\put(1772,5714){\makebox(0,0)[lb]{$z$}}
\put(3542,3614){\makebox(0,0)[lb]{$\Pi$}}
\put(722,4949){\makebox(0,0)[lb]{$\pth$}}
\put(1550,2864){\makebox(0,0)[lb]{$\Pi(\pth)$}}
\end{picture}
}
\end{center}
\caption{A schematic diagram of a spin-half particle undergoing a cyclic 
state evolution.  Points on 
the sphere correspond to physically distinguishable states.  Going through
each point is a ray, on which phase information and normalisation is 
recorded.} \label{cyclic}
\end{figure}
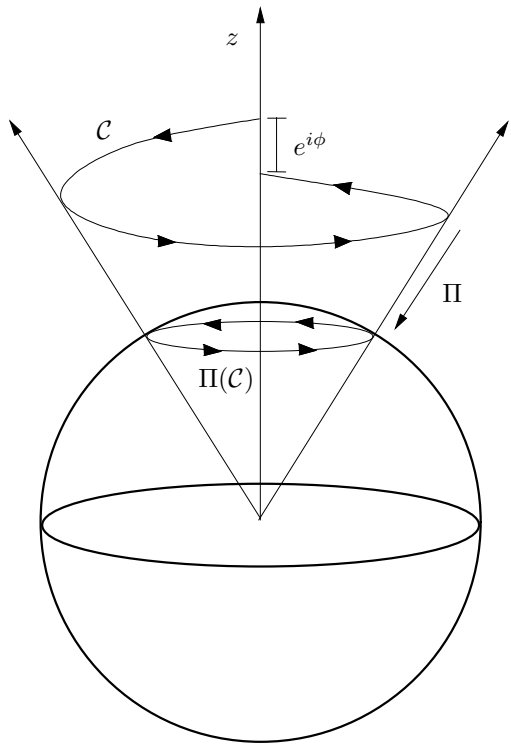

If a system undergoes a cyclic evolution, the ket representing the
system state traces out a path,
$\pth:[0,\tau]\longrightarrow\hilb$, where $\Pi(\pth)$ is a closed
curve in $\prj$, as illustrated in Figure~\ref{cyclic}. In other
words, the initial and final states should be on the same ray in
$\hilb$, but may be related by a phase, $e^{i\phi}$.  We will
measure this phase with respect to a reference curve in $\hilb$:
for each point $\ket{\psi(t)}$ on $\pth$, we can choose a smoothly
varying representative $\ket{{\tilde\psi(t)}}$ from $\Pi(\psi(t))$
in such a way that $\ket{{\tilde\psi(0)}} =
\ket{{\tilde\psi(\tau)}}$. We can then write
\begin{equation}
\ket{\psi(t)}=e^{if(t)}\ket{\tilde\psi(t)}
\label{vectorfield}
\end{equation}
so that the phase
change of $\ket{\psi(0)}$ associated with the cyclic evolution is
given by $\phi=f(\tau)-f(0)$.

\subsection{Dynamic and geometric phase}
The time evolution of a quantum system is governed by the
Sch\"odinger equation,
\begin{equation}
i\hbar\frac{\d}{\d t}\ket{\psi(t)}=H(t)\ket{\psi(t)},
\end{equation}
where $H(t)$ is the Hamiltonian. Substituting
Eq.~(\ref{vectorfield}) into the above, rearranging and
multiplying by $\bra{\psi(t)}$ gives the following~\cite{Ana92},
\begin{equation}
\frac{\d f(t)}{\d t}=
-\frac{1}{\hbar}\bra{\psi(t)}H\ket{\psi(t)} +
i\bra{\tilde\psi(t)}\frac{\d}{\d t}\ket{\tilde\psi(t)},
\end{equation}
or, when integrated,
\begin{equation}
\phi=
-\frac{1}{\hbar}\int_{0}^{\tau}\bra{\psi(t)}H\ket{\psi(t)}\d t
+i\int_{0}^{\tau}\bra{\tilde{\psi}(t)}\frac{\d}{\d t}\ket{\tilde{\psi}(t)} \d t.
\label{totalphase}
\end{equation}
Thus, $\phi$ can be decomposed into a dynamical phase
\begin{equation}
\delta=-\frac{1}{\hbar}\int_{0}^{\tau}\bra{\psi(t)}H\ket{\psi(t)}\d t
\label{dphase}
\end{equation}
which depends on the Hamiltonian, and a geometric phase
\begin{equation}
\gamma=i\oint_{\mathcal{C}}\bra{\tilde\psi}\d\ket{\tilde\psi}
\label{gphase}
\end{equation}
which depends only on the path $\pth$; $\gamma$ is independent
of the rate at which
$\ket{\psi(t)}$ progresses along $\pth$, the Hamiltonian, or the
choice of reference $\left\{\ket{\tilde\psi}\right\}$.

\subsection{Berry's phase}
A particular instance of this geometric phase is Berry's
phase~\cite{Ber84}, which occurs when the adiabatic theorem (see
\cite{Galindo90}) is satisfied. In this case, if the initial state
$\ket{\psi(0)}$ of the system is an eigenstate of the Hamiltonian,
the state $\ket{\psi(t)}$ remains an eigenstate
$\ket{\psi(t)}=\ket{n(\vec{R})}$ of the instantaneous Hamiltonian
$H({\bf R})$, where $\vec{R}$ is a set of time-varying parameters
controlling the Hamiltonian. Supposing $\vec{R}$ traces a closed
loop in the parameter space, the geometric phase of the system can
be written in terms of $\vec{R}$.  In this case, provided the
energy eigenspace of the instantaneous Hamiltonians is
non-degenerate along the path $\mathcal C$, the geometric phase
acquired by the $n^{\rm{th}}$-eigenstate is
\begin{equation}
\gamma_{n}=i\oint_{C}\bra{\widetilde{n({\bf R})}}\nabla_{\bf R}\ket{\widetilde{n({\bf
R})}}\cdot d{\bf R},
\end{equation}
where $\nabla_{\bf R}$ is the gradient operator with respect to the
parameters $\vec{R}$, and $\ket{\widetilde{n({\bf R})}}$
is defined as in the previous section. This line integral can be
transformed into a surface integral over any surface in the
parameter space whose boundary is $\pth$.

Since experimentally it is much easier to control the Hamiltonian than
the actual state of a system, the adiabatic
case is of importance. However, the adiabatic conditions
necessarily mean that the processes take a long time compared to
the characteristic dynamical time-scales, and thus are much slower
than dynamic methods of generating phases.

\section{Single-qubit evolution} \label{sQubit}

\subsection{Qubit Dynamics} \label{ssDynamics}
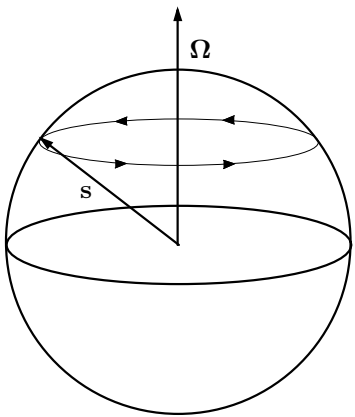
\begin{figure}
\begin{center}
\setlength{\unitlength}{0.00043333in}
\begingroup\makeatletter\ifx\SetFigFont\undefined%
\gdef\SetFigFont#1#2#3#4#5{%
  \reset@font\fontsize{#1}{#2pt}%
  \fontfamily{#3}\fontseries{#4}\fontshape{#5}%
  \selectfont}%
\fi\endgroup%
{
\begin{picture}(4265,5005)(0,-10)
\put(2400,4600){\makebox(0,0)[t]{$\vec{\Omega}$}}
\put(1000,2800){\makebox(0,0)[t]{$\vec{s}$}}
\thicklines
\put(2135,2088){\ellipse{4230}{960}}
\put(2127,2126){\ellipse{4224}{4224}}
\thinlines
\put(2135,3348){\ellipse{3420}{570}}
\thicklines
\path(2120,2088)(2120,4968)
\blacken\path(2157.500,4848.000)(2120.000,4968.000)(2082.500,4848.000)(2157.500,4848.000)
\path(2135,2088)(455,3388)
\blacken\path(572.854,3344.220)(455.000,3388.000)(526.955,3284.904)(572.854,3344.220)
\thinlines
\blacken\path(1500,3558)(1365,3603)(1500,3658)
\path(1500,3558)(1365,3603)(1500,3658)
\blacken\path(1355,3133)(1490,3088)(1355,3033)
\path(1355,3133)(1490,3088)(1355,3033)
\blacken\path(2815,3573)(2680,3618)(2815,3673)
\path(2815,3573)(2680,3618)(2815,3673)
\blacken\path(2680,3128)(2815,3083)(2680,3028)
\path(2680,3128)(2815,3083)(2680,3028)
\end{picture}
}

\end{center}
\caption{Solution to the equations of motion for a single spin-half
particle.} \label{fDynamics}
\end{figure}

Here we will focus on developing an understanding
of the time evolution of a single qubit governed by a very
general Hamiltonian.
Recall that any $2\times 2$ Hermitian matrix can be written in
terms of the unit matrix
and the three Pauli matrices. In particular a single qubit density
operator can be parametrised as
\begin{equation}
\rho=\half(\Id+\vec{s}\cdot\sigma)=\half\left(
\begin{array}{cc}
1+s_z & s_x-i s_y\\ s_x+i s_y& 1-s_z
\end{array}
\right),
\label{Bloch}
\end{equation}
where the real vector $\vec{s}=(s_x, s_y, s_z)$ is known as
the Bloch vector. By the same token any $2\times 2$ Hamiltonian can be written
as
\begin{equation}
H=\frac{\hbar}{2}(\Omega_0 \Id+{\bf \Omega}\cdot\sigma) ,
\label{Rabi}
\end{equation}
where ${\bf \Omega}$ is called the Rabi vector. Substituting
Eq.~(\ref{Bloch}) and Eq.~(\ref{Rabi}) into the equation of motion
for the density operator,
\begin{equation}
i \hbar \frac{\d}{\d t} \rho = [H, \rho]
\end{equation}
and using the identity
\begin{equation}
({\bf a}\cdot\sigma)({\bf b}\cdot\sigma)= ({\bf a}\cdot {\bf b})\;
\Id +i ({\bf a}\times {\bf b})\cdot \sigma ,
\end{equation}
we find the following equation of motion for the Bloch
vector,
\begin{equation}
\frac{\d}{\d t}\; {\bf s} = {\bf \Omega}\times {\bf s}.
\label{Dynamics}
\end{equation}
This equation has a simple
geometric solution: vector ${\bf s}$ revolves around vector ${\bf
\Omega}$ with angular frequency given by $|{\bf
\Omega}|$, the length of ${\bf \Omega}$, as illustrated in
Figure \ref{fDynamics}.

From Eq.~(\ref{Dynamics}), it is relatively easy to move to
situations typical of those encountered in quantum computation.
Hamiltonians that describe qubits interacting with external
potentials are usually time dependent. Typical external
perturbations are periodic such as, for example, spins coupled to
oscillating magnetic fields in NMR or atomic dipole moments
coupled to oscillating electromagnetic field in the optical
domain.  Within the Rotating Wave Approximation the oscillating
field can be replaced by a rotating field, and so the Hamiltonian
is of the form

\begin{equation}
H(t)=\frac{\hbar}{2}\left(
\begin{array}{cc}
\omega_0 & \omega_1 e^{-i(\omega t +\phi)}\\ \omega_1 e^{i(\omega t +\phi)}& -\omega_0
\end{array}
\right),
\label{TypicalH}
\end{equation}
where $\omega_0/2\pi$ is the system's transition frequency,
while $\omega/2\pi$ and $\hbar\omega_1$ are the frequency and the
amplitude of the oscillating field, respectively. This gives
\begin{equation}
\Omega_x = \omega_1\cos(\omega t +\phi),\quad
\quad \Omega_y = \omega_1\sin(\omega t +\phi),\quad\quad
\Omega_z=\omega_0.
\end{equation}

In order to solve Eq.~(\ref{Dynamics}) it is convenient to
consider the evolution of ${\bf s}$ in a frame which rotates with
frequency $\omega$ around the $z$-axis. More precisely, we write
\begin{equation}
{\bf s}(t)=R_z(\omega t){\bf s'}(t),\quad  \quad {\bf
\Omega}(t)=R_z(\omega t){\bf \Omega'}(t),
\label{rotating}
\end{equation}
where $R_z(\omega t)$ is the rotation matrix
\begin{equation}
R_z(\omega t)=
\left(
\begin{array}{ccc}
\cos(\omega t) & -\sin(\omega t) & 0\\
\sin(\omega t) & \cos(\omega t) & 0\\
0 & 0 & 1
\end{array}
\right)
 = \exp (\omega t M_z),
\end{equation}
for
\begin{equation}
M_z=\left(
\begin{array}{ccc}
0 & -1& 0\\ 1 & 0 & 0\\ 0 & 0 & 0
\end{array}
\right).
\end{equation}
Substituting Eq.~(\ref{rotating}) into Eq.~(\ref{Dynamics}) and
taking into account that
\begin{equation}
\frac{\d}{\d t}\; R_z(\omega t)= R_z(\omega t)(\omega M_z),\quad\quad
M_z {\bf s}' = \hat{\vec{z}} \times {\bf s}',
\end{equation}
where $\hat{\vec{z}}$ is a unit vector in the $z$-direction, we
obtain
\begin{equation}
\frac{\d}{\d t}\; {\bf s'} = {\bf \Omega'}\times {\bf s'}
\label{rotDynamics}
\end{equation}
with the time-independent vector ${\bf \Omega'}$,
\begin{equation} \label{eVecOmegaPrime}
\Omega'_x = \omega_1\cos(\phi),\quad\quad
\Omega'_y = \omega_1\sin(\phi),\quad\quad
\Omega'_z=\omega_0-\omega.
\end{equation}

If we can control the strength of coupling $\hbar\omega_1$, the
frequency $\omega$ and the phase $\phi$ of the external field we
can prepare any vector ${\bf \Omega}$. This implies that if we
know the initial state of the qubit then with a single rotation we
can position the Bloch vector ${\bf s}$ in any prescribed
direction.

\subsection{Calculating geometric phases}\label{sCgp}
We can now apply the results of the previous section to the task
of developing a geometric phase of a spin-half particle located
in an external oscillating field.
By varying the parameters of the Hamiltonian adiabatically we will
send a qubit through a cyclic evolution, whose associated
geometric phase can be calculated using Eq.~(\ref{gphase}).

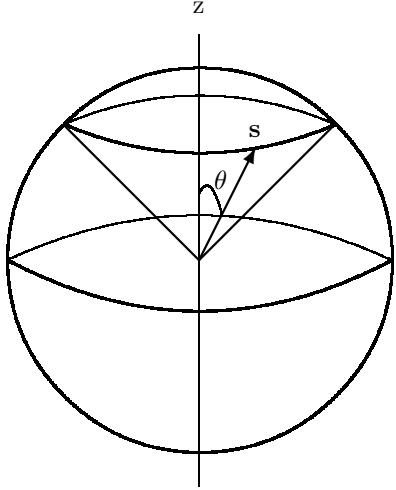
\begin{figure}
\setlength{\unitlength}{1.5mm}
\begin{center}
\begin{picture}(34.1,45)(0,-3)
    \thicklines
    \qbezier (5,5)(0,10)(0,17.05)
    \qbezier (0,17.05)(0,24.1)(5,29.1)
    \qbezier (5,29.1)(10,34.1)(17.05,34.1)
    \qbezier (17.05,34.1)(24.1,34.1)(29.1,29.1)
    \qbezier (29.1,29.1)(34.1,24.1)(34.1,17.05)
    \qbezier (34.1,17.05)(34.1,10)(29.1,5)
    \qbezier (29.1,5)(24.1,0)(17.05,0)
    \qbezier (17.05,0)(10,0)(5,5)
    \qbezier (5,29.1)(17.05,24)(29.1,29.1)
    \qbezier (0,17.05)(17.05,8)(34.05,17.05)
    \put(17.05,17.05){\line(1,1){12.05}}
    \put(17.05,17.05){\line(-1,1){12.05}}
    \put(17.05,17.05){\vector(1,2){5}}
    \thinlines
    \put(17.05,-3){\line(0,1){40}}
    \qbezier (5,29.1)(17.05,34.2)(29.1,29.1)
    \qbezier (17.05,23)(18,25)(19.05,21.05)
    \qbezier (0,17.05)(17.05,25)(34.05,17.05)
    \put(17.05,40){\makebox(0,0)[t]{z}}
    \put(22,29){\makebox(0,0)[t]{$\mathbf{s}$}}
    \put(19,25){\makebox(0,0)[t]{$\theta$}}
\end{picture}
\end{center}
\caption{Spin-half particle in a magnetic field} \label{spinhalf}
\end{figure}

Working in the rotating frame, the components
$(\Omega_x',\Omega_y',\Omega_z')$ of the Hamiltonian are given by
Eq.~(\ref{eVecOmegaPrime}), where the frequency and power of the
rotating field can be used to control the value of the angle
$\theta$ of Figure \ref{spinhalf}.  In the absence of a rotating
field the Rabi vector lies along the $z$-axis, and as the power of
the rotating field is slowly increased from zero to $\omega_1$ the
vector tilts towards the $xy$-plane.  If the Bloch vector
$\vec{s}'$ is initially aligned with $\vec{\Omega}'$ then by the
adiabatic theorem, it will remain aligned with $\vec{\Omega}'$
provided $\vec{\Omega}'$ varies slowly. Therefore, from the
relation between the different components $\vec{\Omega'}$ we find
that the angle $\theta$ between the Bloch vector and the $z$-axis
will be
\begin{equation}\label{costheta}
\cos\theta=\frac{\Omega'_z}{\sqrt{\Omega^{'2}_x+\Omega_y^{'2}+\Omega^{'2}_z}}
=\frac{\omega_0-\omega}{\sqrt{(\omega_0-\omega)^2+\omega_1^2}}.
\end{equation}

Varying the phase $\phi$ of the rotating field from
Eq.~(\ref{eVecOmegaPrime}) will then cause $\vec{s}'$ to rotate
around the $z$-axis.  The geometric phase associated with a full
revolution is easily calculated by parameterizing the state in
terms of the $\upket$ and $\downket$ eigenstates of quantization
along the $z$-axis,
$\ket{\tilde\psi(\alpha)}=\cos(\theta/2)\upket+\sin(\theta/2)e^{i\alpha}\downket$,
where $\alpha$ changes smoothly from $0$ to $2\pi$. Using
Eq.~(\ref{gphase}),
\begin{eqnarray} \label{ePhaseHalf}
\gamma & = &
i\oint_{\mathcal{C}}\left(\cos\frac{\theta}{2}\upket+\sin\frac{\theta}{2}e^{i\alpha}\downket\right)^\dagger
\d\left(\cos\frac{\theta}{2}\upket+\sin\frac{\theta}{2}e^{i\alpha}\downket\right)\nonumber\\
       & = &
- \int_{0}^{2\pi}\left(\cos\frac{\theta}{2}\upbra+\sin\frac{\theta}{2}e^{-i\alpha}\downbra\right)
\left(\sin\frac{\theta}{2}e^{i\alpha}\downket\right)\d\alpha\nonumber\\
       & = &
-\frac{1}{2}(1-\cos\theta)\int_{0}^{2\pi}\d\alpha\nonumber\\
       & = &
-\pi(1-\cos\theta).
\end{eqnarray}
Furthermore, the result can be generalized to any closed path with
the result that the geometric phase is equal to half the solid
angle enclosed by $\pth$ on the Bloch sphere \cite{Ber84}.

\subsection{Eliminating dynamic phases} \label{sElimination}
In order to perform conditional quantum gate operations using
geometric phases only, it is necessary to find a way to eliminate
the dynamic phase. One approach is to use a refocussing technique known as
spin-echo. The basic idea is to apply the cyclic
evolution twice, with the second application surrounded by a pair
of fast $\pi$ transformations (this being simply the
transformation that swaps the basis states $\upket$ and
$\downket$.) The net effect of this compound transformation would
be to cancel \emph{all} the acquired phases except that the second
cyclic evolution is performed by retracing the first but in the
opposite direction so that while the dynamic phases cancel, the
geometric phases do not.

To see why this is so, let $\pth_\ua$ be the closed curve in $\prj$
traced out by $\ket{\tilde{\uparrow}}$ during the first cyclic evolution
and $\pth_\da$ the one traced out by
$\ket{\tilde{\downarrow}}$, with corresponding dynamic and geometric phases
$\delta_\ua$, $\gamma_\ua$, $\delta_\da$, and $\gamma_\da$.  Referring back
to Eq.~(\ref{ePhaseHalf}) we see that $\gamma_\ua=-\gamma$ and $\gamma_\da=\gamma$
for $\gamma = \pi(1-\cos\theta)$.
Similarly, if we write $\bar{\pth}_\ua$ and $\bar{\pth}_\da$ for
the second cyclic evolution, these are simply
$\pth_\ua$ and $\pth_\da$ carried out in opposite orientations so
they have corresponding phases $\bar{\gamma}_\ua = \gamma$ and
$\bar{\gamma}_\da = -\gamma$.

In summary, we can follow the states through the compound evolution
as follows:
\setlength{\arraycolsep}{0mm}
\begin{equation}
\begin{array}{ccccc}
\upket
& \stackrel{\pth_\ua}{\longrightarrow}
  e^{i(\delta_\ua - \gamma)} \upket
& \stackrel{\pi}{\longrightarrow}
  e^{i(\delta_\ua - \gamma)} \downket
& \stackrel{\bar{\pth}_\da}{\longrightarrow}
  e^{i(\delta_\ua + \delta_\da - 2\gamma)} \downket
& \stackrel{\pi}{\longrightarrow}
  e^{i(\delta_\ua + \delta_\da - 2\gamma)} \upket \\
\downket
& \stackrel{\pth_\da}{\longrightarrow}
  e^{i(\delta_\da + \gamma)} \downket
& \stackrel{\pi}{\longrightarrow}
  e^{i(\delta_\da + \gamma)} \upket
& \stackrel{\bar{\pth}_\ua}{\longrightarrow}
  e^{i(\delta_\ua + \delta_\da + 2\gamma)} \upket
& \stackrel{\pi}{\longrightarrow}
  e^{i(\delta_\ua + \delta_\da + 2\gamma)} \downket.
\end{array}
\end{equation}
\setlength{\arraycolsep}{1mm} Since the global phase factor
$e^{i(\delta_\ua + \delta_\da)}$ is not physical this sequence of
operations behaves as promised.  The dynamic phases are eliminated
and we are left with an exclusively geometric phase difference of
$4\gamma = 4\pi \cos\theta$.

\section{Conditional dynamics} \label{sConditional}

\subsection{2-Spin Hamiltonian}
This geometric phase can be used to implement a 2-qubit controlled-phase
gate. Consider to begin with a system of two non-interacting spin-half
particles $S_a$ and $S_b$.
In a reference frame aligned with the static field, the Hamiltonian reads
\begin{equation}
H_0 = \hbar \omega_a S_{az}\otimes \Id_b + \hbar \omega_b \Id_a\otimes S_{bz},
\end{equation}
or, in the basis $\{ \ket{S_{az}, S_{bz}}\}_{S_{az}, S_{bz}} = \{ \ket{\uparrow \uparrow}, \ket{\uparrow \downarrow}, \ket{\downarrow\uparrow}, \ket{\downarrow\downarrow}\}$,
\begin{equation}
H_0 = \frac{\hbar}{2}\left(\begin{array}{cccc} \omega_a+\omega_b & 0 & 0 & 0 \\ 0 & \omega_a-\omega_b & 0 & 0 \\ 0 & 0 & -\omega_a+\omega_b & 0 \\ 0 & 0 & 0 & -\omega_a -\omega_b \end{array}\right),
\end{equation}
where the frequencies $\omega_a/2\pi$ and $\omega_b/2 \pi$ are the
transition frequencies of the two spins and we have used the scaled
Pauli operators $S_i = \sigma_i/2$.
(From now on we assume that $\omega_a$ and $\omega_b$ are very different with
$\omega_a > \omega_b$.)

If the two particles are sufficiently close to each other, they
will interact, creating additional splittings between the energy
levels.  In the case of  two spin-half particles, the
magnetic field of one spin may directly or indirectly affect the 
energy levels of the
other spin; the energy of the system is increased by $\pi \hbar
J/2$ if the spins are parallel and decreased by $\pi \hbar J/2$ if
the spins are antiparallel. The Hamiltonian of the system taking
into account this interaction reads
\begin{equation}
H=H_0+2\pi\hbar J S_{az}\otimes S_{bz},
\end{equation}
or, in the previously chosen basis,
\begin{equation}
H=\frac{\hbar}{2}\left(\begin{array}{cccc} \omega_a+\omega_b+\pi J & 0 & 0 & 0 \\ 0 & \omega_a-\omega_b-\pi J & 0 & 0 \\ 0 & 0 & -\omega_a+\omega_b-\pi J & 0 \\ 0 & 0 & 0 & -\omega_a -\omega_b+\pi J \end{array}\right).
\end{equation}

\begin{figure}
\begin{center}
\setlength{\unitlength}{0.03in}
\begin{picture}(100,80)
\put(10,18){$\ket{\downarrow \downarrow}$}
\put(23,18){$\frac{-\omega_a-\omega_b}{2}$}
\put(40,20){\line(1,0){20}} \put(60,20){\line(2,1){5}}
\put(65,22.5){\line(1,0){20}}
\put(90,20.5){$\frac{-\omega_a-\omega_b+\pi J}{2}$}

\put(10,33){$\ket{\downarrow \uparrow}$}
\put(23,33){$\frac{-\omega_a+\omega_b}{2}$}
\put(40,35){\line(1,0){20}}
\put(60,35){\line(2,-1){5}}
\put(65,32.5){\line(1,0){20}}
\put(90,30.5){$\frac{-\omega_a+\omega_b-\pi J}{2}$}

\put(10,53){$\ket{\uparrow \downarrow}$}
\put(25,53){$\frac{\omega_a-\omega_b}{2}$}
\put(40,55){\line(1,0){20}}
\put(60,55){\line(2,-1){5}}
\put(65,52.5){\line(1,0){20}}
\put(90,50.5){$\frac{\omega_a-\omega_b-\pi J}{2}$}

\put(10,68){$\ket{\uparrow \uparrow}$}
\put(25,68){$\frac{\omega_a+\omega_b}{2}$}
\put(40,70){\line(1,0){20}}
\put(60,70){\line(2,1){5}}
\put(65,72.5){\line(1,0){20}}
\put(90,70.5){$\frac{\omega_a+\omega_b+\pi J}{2}$}

\put(70,32.5){\vector(0,1){40}}
\put(73,60){$\omega_+$}

\put(80,22.5){\vector(0,1){30}}
\put(83,40){$\omega_-$}
\end{picture}
\end{center}
\caption{The energy diagram of two interacting spin-half nuclei. The transition frequency of the first spin depends on the state of the second spin.}\label{energydiagram}
\end{figure}
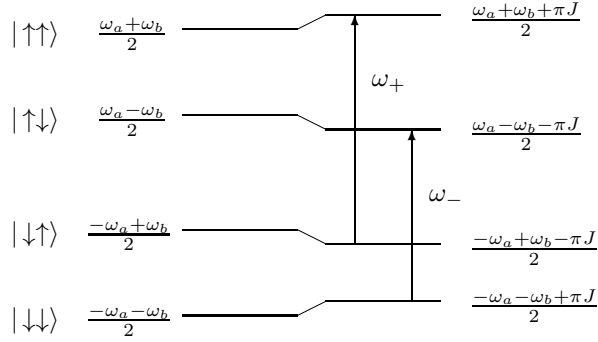

Figure~\ref{energydiagram} illustrates the energy levels of the
system. When spin $S_b$ is in state $\ket{\uparrow}$, the
transition frequency of the spin $S_a$ is
\begin{equation}
\omega_+ = \omega_a + \pi J,
\end{equation}
whereas when spin $S_b$ is in state $\ket{\downarrow}$, the transition frequency of the spin $S_a$ is
\begin{equation}
\omega_- = \omega_a - \pi J.
\end{equation}

\subsection{Conditional phase shift}
Now suppose that in addition to the static field, we apply a rotating field
that is slowly varied as described in Section~\ref{sCgp}.
We have seen that the Berry phase acquired by a spin depends on its
transition resonance frequency as given by Eq.~(\ref{costheta}).
Therefore, at the
end of a cyclic evolution, the Berry phase acquired by the spin $S_a$ will
be different for the two possible states of spin $S_b$. Indeed, when
spin $S_b$ is in state $\ket{\uparrow}$, the Berry phase acquired by the
spin $S_a$ is $\gamma_+=\mp\pi(1 - \cos \theta_+)$, with the sign negative
or positive depending on whether spin $S_a$ is up or down, respectively,
and 
\begin{equation}
\cos \theta_+ = \frac{\omega_+-\omega}{\sqrt{(\omega_+-\omega)^2+\omega_1^2}}.
\end{equation}
Similarly, when spin $S_b$ is in state  $\ket{\downarrow}$, the Berry
phase acquired by the spin $S_a$ is $\gamma_-=\mp\pi(1-\cos \theta_-)$ where
\begin{equation}
\cos \theta_- = \frac{\omega_--\omega}{\sqrt{(\omega_--\omega)^2+\omega_1^2}}.
\end{equation}
As in the single-particle case, it is necessary to eliminate the
dynamic phase in order to construct a purely geometric conditional
phase gate.  
This can be accomplished using almost the same technique as
in the single-particle case described in Section \ref{sElimination}.
In this case, however, we must apply the sequence of operations
\begin{equation}
\pth \longrightarrow 
\pi_a \longrightarrow 
\bar{\pth} \longrightarrow
\pi_b \longrightarrow
\pth \longrightarrow
\pi_a \longrightarrow
\bar{\pth} \longrightarrow
\pi_b,
\end{equation}
where $\pi_a$ and $\pi_b$ are $\pi$-pulses applied to particles
$a$ and $b$, respectively, and $\pth$ and $\bar{\pth}$ are adiabatic
transformations as in Section \ref{sElimination}.
If we define the differential Berry phase shift
\begin{equation}
\Delta\gamma = \gamma_+ - \gamma_- = \pi \left(  \frac{\omega_+-\omega}{\sqrt{(\omega_+-\omega)^2+\omega_1^2}}-\frac{\omega_--\omega}{\sqrt{(\omega_--\omega)^2+\omega_1^2}}\right)
\end{equation}
then the net transformation, up to global phases, is given by
\begin{equation}
\left(\begin{array}{cccc}
e^{2i\Delta\gamma} & 0             & 0             & 0 \\
0            & e^{-2i\Delta\gamma} & 0             & 0 \\
0            & 0             & e^{-2i\Delta\gamma} & 0 \\
0            & 0             & 0             & e^{2i\Delta\gamma}
\end{array}\right).
\end{equation}
Thus, we have succeeded in engineering a conditional evolution since
the state of the qubit $S_b$ influences the phase acquired by a
second qubit $S_a$.\cite{URL}  
This gate, which introduces a phase of $e^{2i\Delta\gamma}$ if the
two spins are aligned and $e^{-2i\Delta\gamma}$ if they are
anti-aligned, is equivalent to the controlled phase gate
introduced in Section \ref{sBasics}~\cite{Jones98c}.

\subsection{Fault Tolerance}
\begin{figure}
\begin{center}
\epsfig{figure=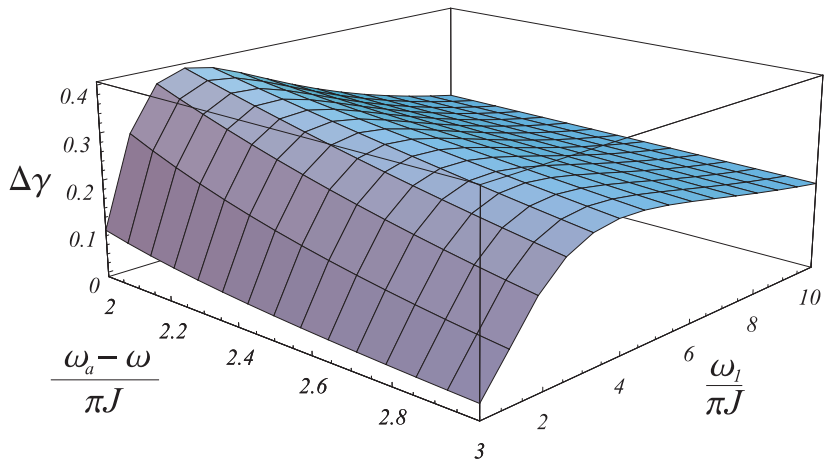,width=.9\textwidth}
\end{center}
\caption{Plot of differential phase shift $\Delta\gamma$ as a function
of $\frac{\omega_a - \omega}{\pi J}$ and $\frac{\omega_1}{\pi J}$.}
\label{gFaultTolerance}
\end{figure}

The form of the dependence of $\Delta\gamma$ on the detuning
$\omega_a - \omega$ and the amplitude of the oscillating field
$\omega_1$ builds into the geometric phase gate a natural type of
fault tolerance not present in the simple non-geometric
conditional phase gate.  In many experiments, such as NMR, it is
easy to control the detuning quite precisely, but relatively
difficult to control the amplitude of the oscillating field.
Figure \ref{gFaultTolerance} plots $\Delta\gamma$ as a function of
these parameters in units of $\pi J$ for a range of values: for
fixed $\omega_a - \omega$, we see that there is a peak in
$\Delta\gamma$ as a function of $\omega_1$. Therefore, if
$\omega_1$ is chosen to coincide with this peak then
$\Delta\gamma$ will be insensitive to errors in $\omega_1$ to
first order.  As the height of the peak depends on the detuning,
any desired controlled phase gate can be achieved.

\section{Conclusions} \label{sConclusions}
The techniques described in this paper constitute a novel approach to
quantum computation, one that builds entangling gates entirely out of
conditional geometric phases.  These techniques are readily implementable
with current technology in quantum optics and have already been
demonstrated by some of the authors using NMR~\cite{Jones00}.
It would be interesting to further analyse the robustness of geometric
quantum computation to errors.  While it has been observed that
geometric phases are robust to certain types of noise in the
classical parameters controlling the Hamiltonian, it has not
been determined how geometric phases behave in the presence of
decoherence or depolarisation of the quantum system.

\section{Acknowledgments}
This work was supported in part by the European TMR Research
Network ERP-4061PL95-1412, The Royal Society of London, Elsag,
Starlab (Riverland NV, Belgium), the European Science Foundation,
CESG, and the Rhodes Trust.


\end{document}